# Analysis of misidentifications in TEM characterization of organic-inorganic hybrid perovskite material

Yu-Hao Deng [1*], Leon Georg Nest [2]

[1] Academy for Advanced Interdisciplinary Studies, Peking University, Beijing, China

[2] Department of Physics, Freie Universität Berlin, Berlin, German

* Correspondence should be addressed to yuhaodeng@pku.edu.cn

**Abstract**

Organic-inorganic hybrid perovskites (OIHPs) have recently emerged as groundbreaking semiconductor materials owing to their remarkable properties. Transmission electron microscopy (TEM), as a very powerful characterization tool, has been widely used in perovskite materials for structural analysis and phase identification. However, the perovskites are highly sensitive to electron beams and easily decompose into $PbX_2$ (X= I, Br, Cl) and metallic Pb. The electron dose of general high-resolution TEM is much higher than the critical dose of $MAPbI_3$, which results in universal misidentifications that $PbI_2$ and Pb are incorrectly labeled as perovskite. The widely existed mistakes have negatively affected the development of perovskite research fields. Here misidentifications of the best-known $MAPbI_3$ perovskite are summarized and corrected, then the causes of mistakes are classified and ascertained. Above all, a solid method for phase identification and practical strategies to reduce the radiation damage for perovskite materials have also been proposed. This review aims to provide the causes of mistakes and avoid misinterpretations in perovskite research fields in the future.

## Introduction

Organic-inorganic hybrid perovskites (OIHPs), which can be synthesized via low-cost solution-based methods, have emerged as groundbreaking semiconductor materials with remarkable performance in various optoelectronic devices such as solar cells, light-emitting diodes (LED), lasers and photodetectors [1-8]. In terms of structural characterization and phase identification of perovskite materials, transmission electron microscopy (TEM) is considered to be the powerful characterization tool and has been widely used in these fields [9]. Unfortunately, the extreme sensitivity of OIHPs to electron beam irradiation inhibits us from obtaining the real structure of perovskite [10-13]. For example, the best-known $MAPbI_3$ perovskite (Fig. 1A, B) begins to decompose into $PbI_2$ and Pb under 150 $eÅ^{-2}$ total dose irradiation [14]. The degradation process shows in Fig. 1C. However, the value of the electron dose in normal high-resolution transmission electron microscopy (HRTEM) is around 800-2000 $eÅ^{-2}$ $s^{-1}$, which is much higher than the critical dose of $MAPbI_3$ perovskite. Due to the negligence of electron beam-sensitive property of perovskite, the decomposition products, such as $PbI_2$, Pb and other intermediates have been widely misidentified as perovskite in TEM characterizations.

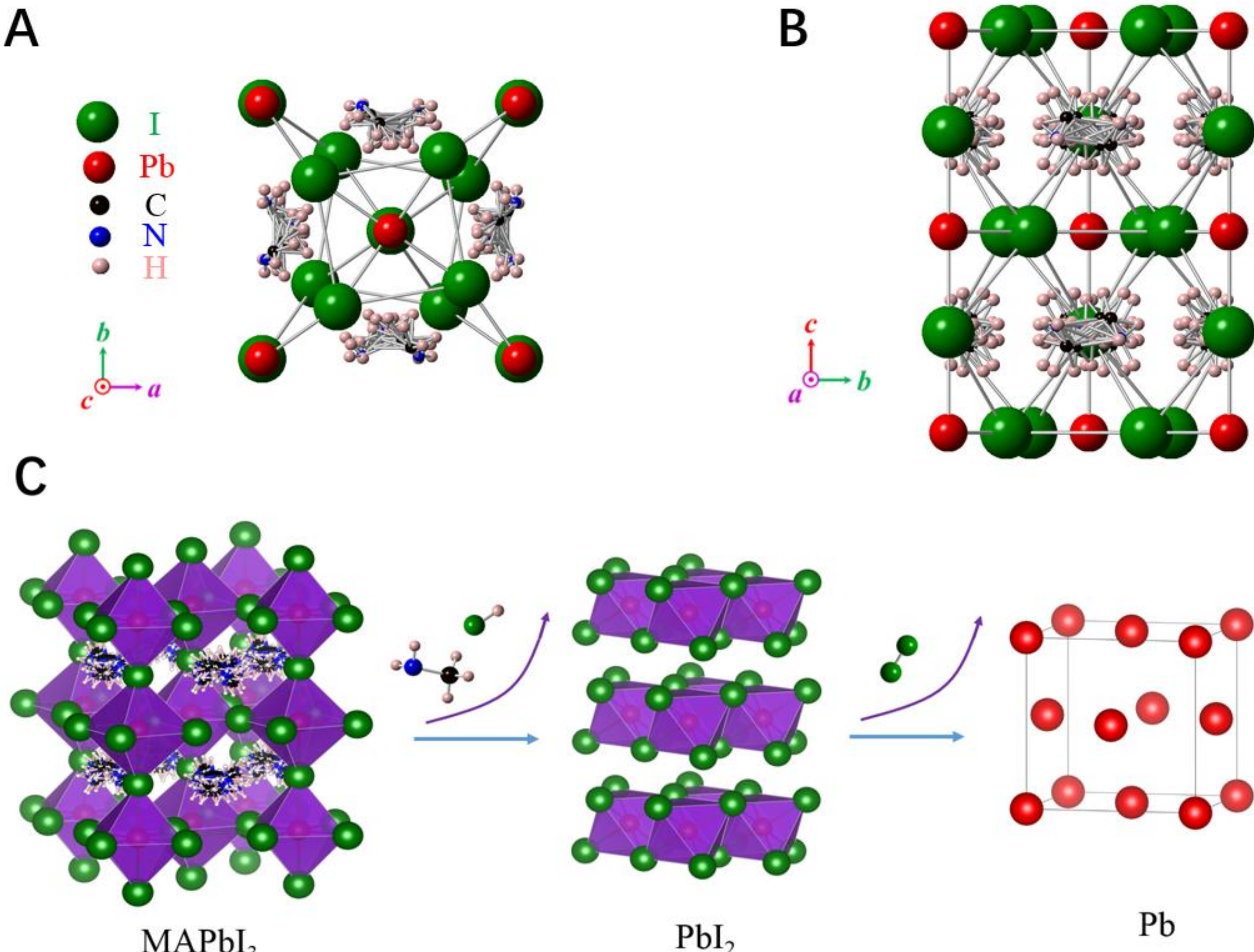

**Fig. 1.** The atomistic structure and degradation process of tetragonal $MAPbI_3$. (A) The atomic models of $MAPbI_3$ along [001] zone axis. (B) The atomic models of $MAPbI_3$ along [100] zone axis. (C) Tetragonal perovskite decomposes into hexagonal $PbI_2$, followed by the escape of methylamine and hydrogen iodide molecules. Furthermore, $PbI_2$ will decompose into cubic metallic Pb by further irradiation of the electron beam. Here, colors represent the following: green, iodine; red, lead; black, carbon; blue, nitrogen; pink, hydrogen.

The widely existed misidentifications have negatively influenced the development of perovskite research field, such as structural determination, material growth, phase transition, heterostructure and so on [6, 15-45]. Although the mistakes are being taken seriously [14, 46], analyzing the causes of misidentifications, and proposing a solid method of phase identification for electron beam-sensitive materials is still highly urgent.

In this review, we focus on the best-known $MAPbI_3$ perovskite and aim to present an overall analysis of existing misidentifications in TEM characterization, other perovskites with similar components can be analyzed with the same approach. Firstly, we will highlight the summary and classification of the mistakes, whose HRTEM images mismatch intrinsic perovskites and electron dose is much higher than the critical dose. Subsequently, analyzing the causes of these misidentifications. The causes are specifically highlighted in this paper and are summarized as ignoring the absent crystal planes, systematic extinction and proofreading of decomposition product. Furthermore, a solid method for phase identification of perovskite materials will be proposed. Finally, the available strategies to obtain the intrinsic structure of perovskites are discussed. This review aims to provide the causes of mistakes and avoid misidentifications in perovskite research fields in the future.

## 1. Ignoring the absent crystal planes

All supposed crystal planes should be present in HRTEM and electron diffraction (ED) pattern for a complete crystal structure. However, the missing crystal planes exist extensively in TEM characterizations in previously published articles. The absence of crystal planes indicates that the structure and composition of perovskite are no longer intrinsic under electron beam irradiation. In contrast to intrinsic $MAPbI_3$ perovskite, the structures of decomposition products were misidentified as perovskite, which we call "pseudo". The "missing crystal planes" in this paper mean "absent reflections" in HRTEM and SAED characterizations rather than particular atomic arrangements in crystals.

Fig. 2A, B show the HRTEM image and Fast Fourier Transform (FFT) of intrinsic $MAPbI_3$ perovskite along [001] zone axis at total doses of 1.5 $e\text{Å}^{-2}$ in room temperature [47]. Obviously, $(1\bar{1}0)$, (110) planes with 0.62 nm interplanar spacing are existing in images, which matches the ED pattern (Fig. 2C) and XRD data of intrinsic $MAPbI_3$ [48, 49]. Fig. 2D, E are the HRTEM image and Fast Fourier Transform (FFT) of the pseudo $MAPbI_3$ perovskite in previously published paper at high total doses under normal TEM condition [16]. $(1\bar{1}0)$, (110) planes are missing and only $(2\bar{2}0)$, (220) planes remains. Actually, the perovskite has decomposed to $PbI_2$, according to the matched ED pattern along $[4\bar{4}1]$ zone axis (Fig. 2F). Similarly, Fig. 2G-I show the HRTEM image, FFT and ED pattern of intrinsic $MAPbI_3$ along $[\bar{2}01]$ zone axis at total doses of 3 eÅ$^{-2}$ in liquid nitrogen temperature [50]. The phase composition in Fig. 2J, K under normal TEM condition is identified as $PbI_2$ rather than $MAPbI_3$ due to the lacking of $(1\bar{1}2)$, (112) planes and the matched ED pattern (Fig. 2L) [43]. The newly added annotations in reproduced HRTEM images were marked by yellow font.

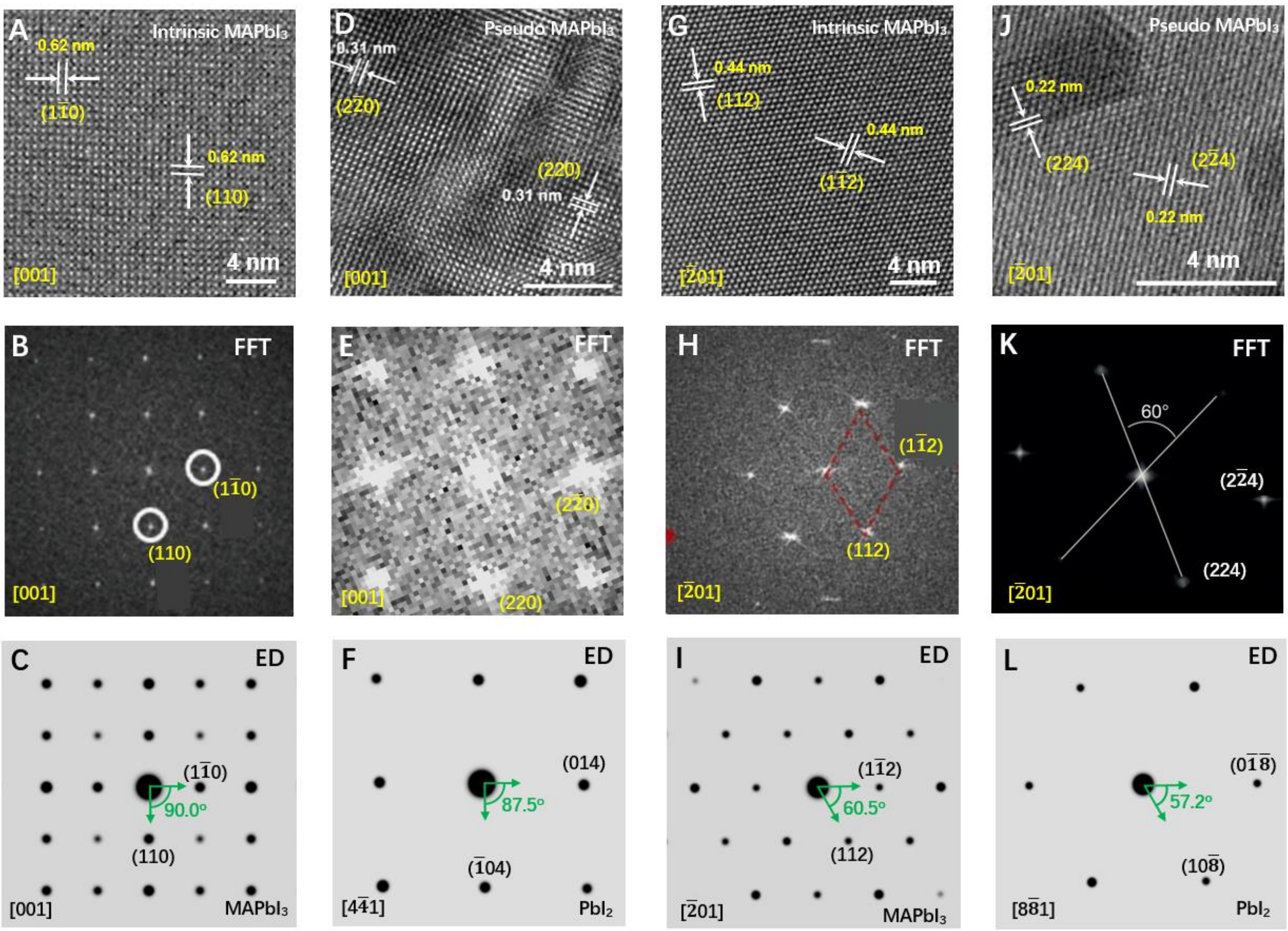

**Fig. 2.** Analysis of ignoring the absent crystal planes. (A) HRTEM image of intrinsic $MAPbI_3$ along [001] axis zone. (B) Fast Fourier Transform (FFT) of Fig. 2A. (C) Simulated ED pattern of intrinsic $MAPbI_3$ along [001] zone axis. (D) HRTEM image of pseudo $MAPbI_3$ along [001] zone axis, $(1\bar{1}0)$, (110) planes are missing. (E) FFT of Fig. 2D. (F) Simulated ED pattern of corrected $PbI_2$ phase along [4$\bar{4}$1] zone axis. Similarly, HRTEM image, FFT and simulated ED patterns of intrinsic (G-I), pseudo (J-L) $MAPbI_3$ along [$\bar{2}$01] zone axis were also be analysed. The newly added annotations in reproduced HRTEM images were marked by yellow font. (A, B) Reproduced with permission from Ref. [47], ©WILEY-VCH Verlag GmbH & Co. KGaA, Weinheim 2020. (D) Reproduced with permission from Ref. [16], ©WILEY-VCH Verlag GmbH & Co. KGaA, Weinheim 2014. (G, H) Reproduced with permission from Ref. [50], © 2020 Elsevier B.V. 2020. (J, K) Reproduced with permission from Ref. [43], © Springer Nature 2015.

Ignoring the absent crystal planes results in the misidentifications. In addition to the [001] and [$\bar{2}$01] axis zones of perovskite, other errors in other zone axes have also been corrected [46]. Even more remarkably, all above intrinsic HRTEM images of $MAPbI_3$ perovskite were captured under low temperature or low-dose electron beam irradiation. To make the comparisons and corrections clearer, Table 1 shows the detailed parameters of the $MAPbI_3$ and $PbI_2$ in Fig. 2.

**Table 1.** Detailed crystallographic parameters of $MAPbI_3$ and $PbI_2$.

| Material and zone axis | Characteristic crystal planes | Interplanar spacing | Interplanar Angle |
| --- | --- | --- | --- |
| **$MAPbI_3$ [001]** | (1$\bar{1}$0), (2$\bar{2}$0)<br>(110), (220) | d(1$\bar{1}$0)= 0.62 nm.<br>d(2$\bar{2}$0)= 0.31 nm.<br>d(110)= 0.62 nm.<br>d(220) =0.31 nm. | <(1$\bar{1}$0), (110)><br>= <(2$\bar{2}$0), (220)> =90.0° |
| **$PbI_2$ [4$\bar{4}$1]** | (014)<br>($\bar{1}$04) | d(014)= 0.32 nm.<br>d($\bar{1}$04) =0.32 nm. | <(014), ($\bar{1}$04)> =87.5° |
| **$MAPbI_3$ [$\bar{2}$01]** | (1$\bar{1}$2), (2$\bar{2}$4)<br>(112), (224) | d(1$\bar{1}$2)= 0.44 nm.<br>d(2$\bar{2}$4)= 0.22 nm.<br>d(112)= 0.44 nm. | <(1$\bar{1}$2), (112)><br>= <(2$\bar{2}$4), (224)> =60.5° |

| | | d(224)= 0.22 nm. | |
|---|---|---|---|
| **$PbI_2$ [$8\bar{8}1$]** | $(0\bar{1}\bar{8})$<br>$(10\bar{8})$ | d$(0\bar{1}\bar{8})$= 0.22 nm.<br>d$(10\bar{8})$= 0.22 nm. | $<(0\bar{1}\bar{8}), (10\bar{8})> =57.2^{o}$ |

## 2. Ignoring the systematic extinction effect

Lattice fringe image and ED pattern in TEM are based on the Bragg's law, described by

$$n\lambda = 2d\cdot\sin\theta \qquad (1)$$

where $n$ is a positive integer, $\lambda$ is the wavelength of the incident wave, $d$ is the interplanar spacing in a crystal and θ is the glancing angle of incidence. Only the crystal planes that satisfy the Bragg diffraction equation can diffract electron beams and appear in TEM image. Besides, some crystal planes satisfying the Bragg's law will still disappear due to the microscopic symmetric elements in crystal structure, such as screw axis, slip plane, centered lattice and so on. We call this phenomenon systematic extinction effect which is often overlooked during phase identification [51].

$MAPbI_3$ perovskite is I4/mcm space group with tetragonal structure, so (100), (011) crystal planes are extinctive and will not appear in TEM image and ED pattern. But in some published papers, extinctive crystal planes appear [36, 52]. As shown in Fig. 3A, B, extinctive {100} crystal planes appear in the SAED images along [001] zone axis. However, the crystal planes should disappear in the intrinsic ED pattern (Fig. 3C). The extra extinctive crystal planes have been demonstrated from the decomposition product $MAPbI_{2.5}$ [14]. Similarly, {011} crystal planes appear in the HRTEM image of pseudo $MAPbI_3$ (Fig. 3D, E) but disappear in the intrinsic ED pattern (Fig. 3F) along [100] zone axis. Moreover, the extinctive (100) and

(011) crystal planes also exist in pseudo perovskite (Fig. 3G, H) but disappear in intrinsic ED pattern (Fig. 3I) along [01$\bar{1}$] zone axis. The materials in Fig. 3D, G cannot be identified as $PbI_2$ or Pb, and more likely to be other intermediates. Ignoring the systematic extinction effect will also lead to misidentifications. In the identification of electron beam-sensitive materials, we should pay more attention to the systematic extinction effect to avoid mistakes.

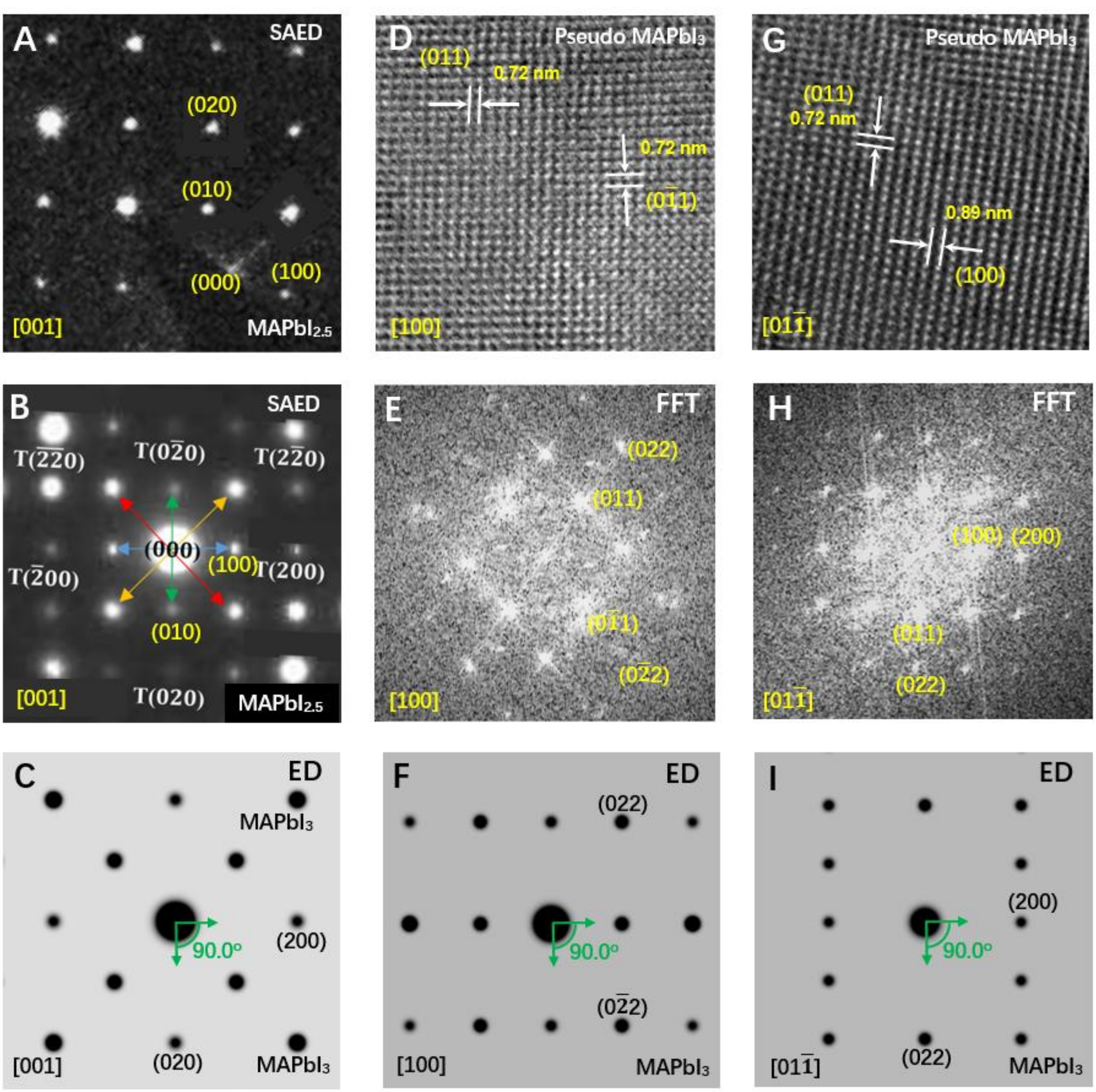

**Fig. 3.** Analysis of ignoring the systematic extinction effect. (A, B) SAED pattern of decomposition product $MAPbI_{2.5}$ along [001] zone axis. (C) Simulated ED pattern of intrinsic $MAPbI_3$ along [001] zone axis. (D) HRTEM image of pseudo $MAPbI_3$ along [100] zone axis. (E) Fast Fourier Transform (FFT) of Fig. 3D. (F) Simulated ED pattern of intrinsic $MAPbI_3$ along [100] axis zone. (G, H) HRTEM image and FFT of pseudo $MAPbI_3$ along [01$\bar{1}$] zone axis. (I) Simulated ED pattern of intrinsic $MAPbI_3$ along [01$\bar{1}$] zone axis. (100),

(010), (011) and (0$\bar{1}$1) crystal planes are extinctive and should not appear in TEM image, FFT and ED pattern. (A) Reproduced with permission from Ref. [14], © Springer Nature 2018. (B) Reproduced with permission from Ref. [36], ©WILEY-VCH Verlag GmbH & Co. KGaA, Weinheim 2018. (D, G) Reproduced with permission from Ref. [52], © Springer Nature 2020.

## 3. Ignoring the proofreading of decomposition product

General phase identification of HRTEM and ED data is to compare a group of interplanar spacings and angles. $MAPbI_3$ perovskite is an electron beam-sensitive material and the critical electron dose is much lower than the radiation level of normal experimental HRTEM and ED. Meanwhile, angles and distances between crystal planes in the $MAPbI_3$ are very similar to the decomposition products, so we need to proofread whether the phase is intrinsic perovskite, $PbI_2$ or Pb. As shown in Fig. 4A [24], the decomposed metallic Pb was incorrectly labeled as perovskite and ED pattern of the corrected phase shows in Fig. 4B. The same misidentification in Fig. 4C and has been corrected in the corresponding ED pattern in Fig. 4D. If the identification is suitable for both intrinsic perovskite and decomposition products, we must refer to the total dose of electron beam radiation. Moreover, it is also not rigorous to only measure one crystal plane to identify the phase, because the decomposed materials have approximate interplanar spacings with intrinsic perovskite. As shown in Fig. 4E-H [19, 22, 24, 33], we completely cannot identify whether these materials are $MAPbI_3$ perovskite, $PbI_2$, Pb or other intermediates based on one crystal plane. It generally seems to happen when the crystal is oriented in a way where lines are formed in the image due to tilting. Ignoring the proofreading of decomposition product makes it impossible to get the reliable result, so rigorous proofreading is necessary and can help us avoid misidentifications of electron beam-sensitive materials.

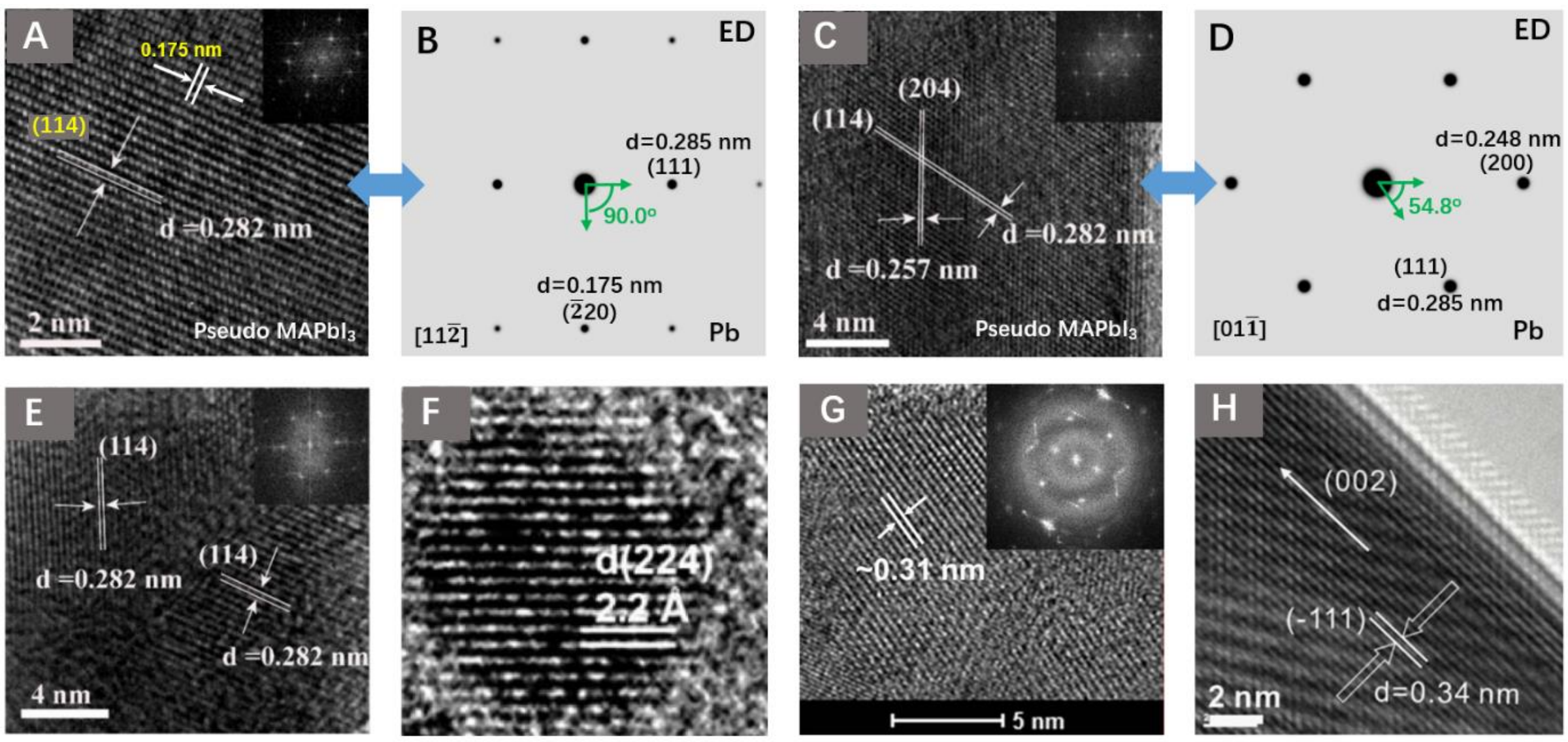

**Fig. 4.** Analysis of ignoring the proofreading of decomposition product. (A) HRTEM image of pseudo $MAPbI_3$. (B) Simulated ED pattern of corrected Pb phase along $[11\bar{2}]$ zone axis. (C) HRTEM image of pseudo $MAPbI_3$. (D) Simulated ED pattern of corrected Pb phase along $[01\bar{1}]$ axis zone. (E-H) HRTEM images of pseudo $MAPbI_3$ with only one crystal plane measured. (A, C, E) Reproduced with permission from Ref. [24], © Multidisciplinary Digital Publishing Institute (MDPI) 2019. (F) Reproduced with permission from Ref. [22], © American Chemical Society 2017. (G) Reproduced with permission from Ref. [19], © Springer Nature 2017. (H) Reproduced with permission from Ref. [33], © IOP Publishing 2014.

## 4. The solid method of phase identification

The misidentifications of perovskite material lead to wrong results, which have negatively affected the development of perovskite research fields. To avoid misleading the HRTEM characterization of perovskite in the future, we propose a solid method of phase identification for electron beam-sensitive materials here. Table 2 shows the crystallographic parameters of intrinsic $MAPbI_3$. Interplanar spacings from experimental data should match the data in the table. The interplanar spacing outside of this table indicates that the phase is not $MAPbI_3$ perovskite and would be other decomposition products. Once a group of alternative crystal planes (($h_1$ $k_1$ $l_1$), ($h_2$ $k_2$ $l_2$)) have successfully corresponded to the intrinsic values, then we need to confirm that the experimental angle between ($h_1$ $k_1$ $l_1$) plane and ($h_2$ $k_2$ $l_2$) plane should also match the intrinsic value. The intrinsic angle between crystal planes can be obtained from

$$\cos\theta = \frac{\frac{h_1 h_2}{a^2}+\frac{k_1 k_2}{b^2}+\frac{l_1 l_2}{c^2}}{\sqrt{\left(\frac{{h_1}^2}{a^2}+\frac{{k_1}^2}{b^2}+\frac{{l_1}^2}{c^2}\right)\left(\frac{{h_2}^2}{a^2}+\frac{{k_2}^2}{b^2}+\frac{{l_2}^2}{c^2}\right)}} \tag{2}$$

where a, b, c are the parameters of the unit cell. Finally, proofread the fast Fourier transform (FFT) of the experimental HRTEM image or experimental ED pattern according to intrinsic ED simulation, only when the experimental data match the ED simulation perfectly can the identification be proved to be successful. Schematic diagram of the method shows in Fig. 5. This method for phase identification is solid and can avoid the problems of ignoring the absent crystal planes, systematic extinction effect and parameter proofreading.

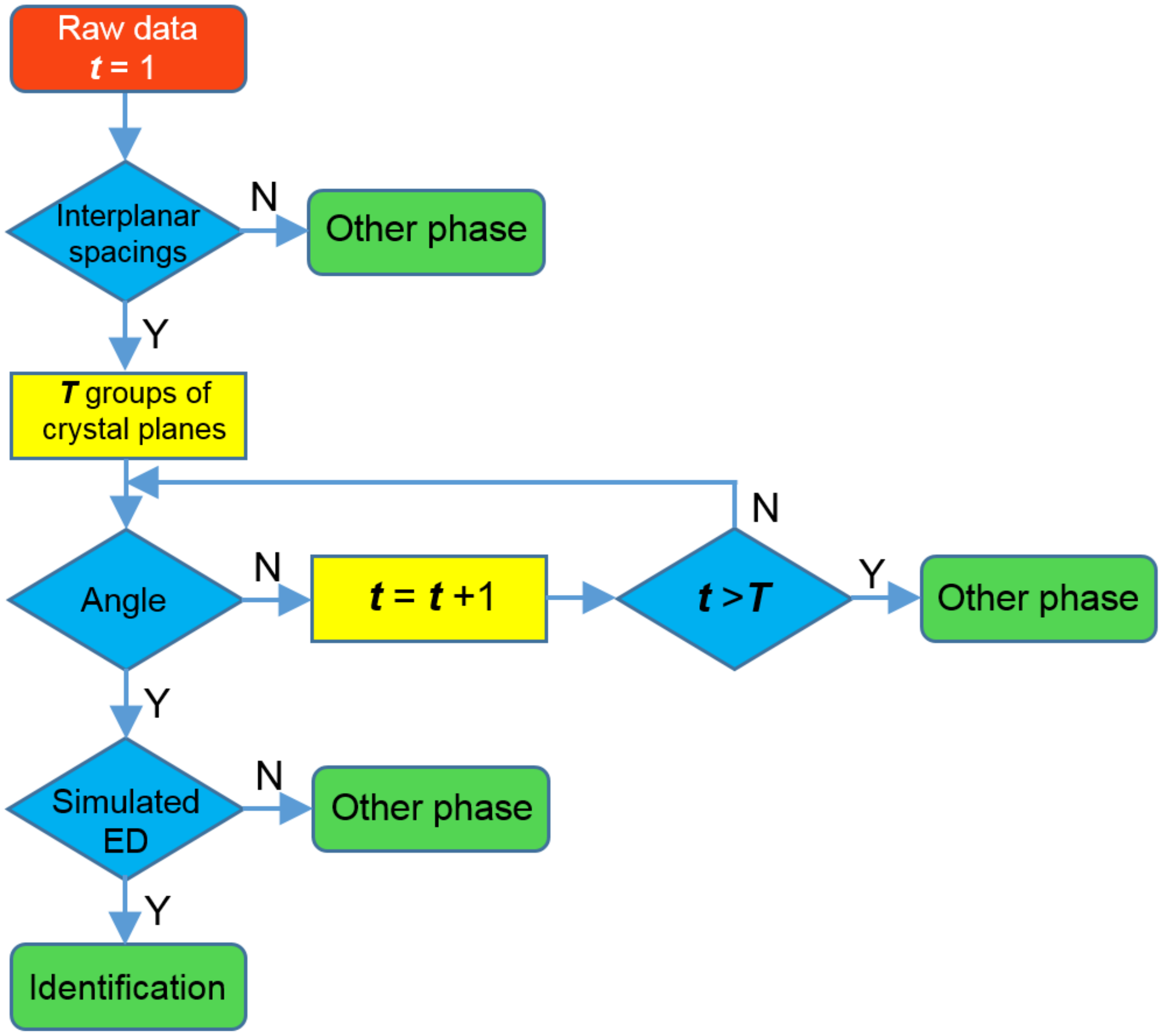

**Fig. 5.** Schematic diagram of the solid method of phase identification. Interplanar spacings and angle between alternative crystal planes must be matched. Moreover, proofreading the fast Fourier transform (FFT) of the experimental HRTEM image or experimental ED pattern according to ED simulation is necessary.

**Table 2.** Crystallographic parameters of tetragonal $MAPbI_3$.

| Sequence number | Index (h k l) | 2θ(°) | Interplanar spacing (Å) | Relative intensity (%) |
|---|---|---|---|---|
| 1 | 002 | 13.9513 | 6.3425 | 57 |
| 2 | 110 | 14.2216 | 6.2225 | 100 |
| 3 | 112 | 19.9730 | 4.4418 | 7 |
| 4 | 200 | 20.1647 | 4.4000 | 10 |
| 5 | 211 | 23.6509 | 3.7587 | 25 |
| 6 | 202 | 24.6041 | 3.6152 | 17 |
| 7 | 004 | 28.1149 | 3.1713 | 43 |
| 8 | 220 | 28.6684 | 3.1113 | 61 |
| 9 | 213 | 31.0176 | 2.8808 | 11 |
| 10 | 114 | 31.6405 | 2.8255 | 30 |
| 11 | 222 | 32.0148 | 2.7933 | 22 |
| 12 | 310 | 32.1387 | 2.7828 | 46 |
| 13 | 204 | 34.8441 | 2.5727 | 6 |
| 14 | 312 | 35.1881 | 2.5483 | 11 |
| 15 | 321 | 37.4940 | 2.3967 | 2 |
| 16 | 224 | 40.5874 | 2.2209 | 38 |
| 17 | 400 | 40.9903 | 2.2000 | 13 |
| 18 | 215 | 42.3526 | 2.1323 | 3 |
| 19 | 323 | 42.6932 | 2.1213 | 1 |

| 20 | 006 | 42.7343 | 2.1142 | 2 |
|---|---|---|---|---|
| 21 | 411 | 42.9354 | 2.1047 | 16 |
| 22 | 314 | 43.2169 | 2.0917 | 34 |
| 23 | 402 | 43.5043 | 2.0785 | 5 |
| 24 | 330 | 43.5998 | 2.0742 | 2 |
| 25 | 116 | 45.2084 | 2.0041 | 1 |

## 5. Strategies to reduce the radiation damage

For most characterizations, and especial on HRTEM, keeping the total dose below the critical value for OIHPs is an extremely challenging job. Herein, we provide several strategies to reduce the damage to electron beam irradiation, and they would be helpful for researchers to obtain the intrinsic structure of perovskites in HRTEM characterizations. Sample protection is a direct way to improve the stability of material [53]. Fan et al. applied hexagonal boron nitride thin films to build an encapsulation layer of $MAPbI_3$, which suggests a much enhanced stability of $MAPbI_3$ and can reduce radiation damage to electron beam [17]. Moreover, keeping imaging at low doses is also an effective method to reduce the radiation damage [14, 54-57]. Direct-detection electron-counting (DDEC) cameras greatly reduce the dose of imaging and enable researchers to capture the intrinsic structure of $MAPbI_3$ at total doses below 3 e Å$^{-2}$ [50, 56]. Zhang *et al*. also invented a program to obtain a direct, one-step alignment of the zone axis and captured HRTEM images automatically at a total dose of 6-12 e Å$^{-2}$ [55]. Carlino developed an in-line holography based on HRTEM system and can get high contrast images at a dose rate of 1-2 eÅ$^{-2}$s$^{-1}$ [58]. Stating the specimen under cryogenic condition can effectively enhance the stability of the material. Using the Cryo-TEM, the atomic structure of $MAPbI_3$ has been captured by HRTEM [50, 59]. In any case, the condition of the intrinsic structure is that the total dose in experimental should be below the critical dose of perovskite.

## 6. Conclusions and outlook

Phase identification plays a vital role in material science, which determines whether we can get correct results from the experimental data [60, 61]. Above summary, classification, correction and analysis of misidentifications in TEM characterization of $MAPbI_3$ perovskite are very helpful for researchers to avoid mistakes in perovskite research fields. The alerts learned from the mistakes and the proposed solid method for phase identification here can not only be applied to OIHPs, but also to other electron beam-sensitive materials, such as metal-organic frameworks (MOFs), organic crystals, etc. In addition, the damage during the process of sample preparation also should be concerned. For example, the widely used focused ion beam (FIB) has been shown to introduce significant damage in the crystal even before imaging takes place [62]. This review provides a sober-minded brain for further TEM characterizations in OIHPs and other electron beam-sensitive materials.

## Methods

Corresponding crystal structures cif files were downloaded from Crystallography Open Database (COD) website. COD IDs of $MAPbI_3$, $PbI_2$ and Pb are 4124388, 9009141 and 9008477 respectively. $MAPbI_3$ is I4/mcm space group with tetragonal structure, cell parameters: a=b=8.839Å, c=12.695Å; α=β=γ=90°. $PbI_2$ is P-3m1 space group with hexagonal structure, cell parameters: a=b=4.555Å, c=20.937Å; α=β=90º, γ=120º. Pb is Fm-3m space group with cubic structure, cell parameters: a=b=c=4.950Å; α=β=γ=90º. The Electron diffraction (ED) simulations of $MAPbI_3$ and $PbI_2$ were obtained using CrystalMaker Software. The interplanar spacing and interplanar angle can be calculated from the cell parameters. During the process of phase identification, I also tried other polytypes of $PbI_2$, but the results did not match well.

**Data availability:** All data are available from the corresponding author(s) upon reasonable request.

**Acknowledgements:** The author thanks the anonymous reviewers for helpful suggestions and editors for communications. The author also appreciates Ying Xiong for suggestions of picture coloring and combination. **This friendly review only aims to perfect the original papers and benefit the development of the academic community**.

**Conflict of interest:** The authors declare no competing financial interest.

**Contributions:** Yu-Hao Deng conceived this work and performed simulations, data analysis and manuscript writing. Leon Georg Nest helped to do the data analysis and paper modification.

**Note added:** This work has been published in ***Journal of Microscopy*** [63].

**Reviewers' reports:** This review paper is timely and an excellent addition to the literature on how to reliably study these materials. Thanks to the author for the editing of the paper, I think the current version is excellent and I'd be happy to see it published; A review paper summarizing and classifying previous mistakes is necessary and will draw more attention from non-experts to this issue. Therefore, I am supportive to this manuscript to be published.